\documentclass[aps,twocolumn,prc,floatfix,showpacs,preprintnumbers,amsmath,amssymb,nofootinbib,groupedaddress]{revtex4}

\usepackage{hyperref}

\usepackage{epsfig,subfigure,latexsym}
\usepackage{rotating}
\usepackage{graphicx}
\usepackage{amsmath}
\usepackage{color}

\usepackage{amsmath,amssymb,amsfonts}
\usepackage{dcolumn}
\usepackage{graphicx}
\usepackage{color}
\allowdisplaybreaks
\begin{document} 

\title{Thermal properties of the nuclear surface}

\author{B. K. Agrawal, D. Bandyopadhyay, J. N. De, and
 S. K. Samaddar} 
\affiliation{
Saha Institute of Nuclear Physics, 1/AF Bidhannagar, Kolkata
{\sl 700064}, India }

\begin{abstract}

 The thermal evolution of a few thermodynamic properties of the
nuclear surface like its thermodynamic potential
energy, entropy and the symmetry free
energy are examined for both semi-infinite nuclear matter and
finite nuclei. The Thomas-Fermi model is employed. Three Skyrme
interactions, namely, SkM$^*$, SLy4 and SK255 are used for the
calculations to gauge the dependence of the nuclear surface properties
on the energy density
functionals. For finite nuclei, the surface observables are computed
from a global liquid-drop inspired fit of the energies and free energies 
of a host of nuclei covering the entire periodic table. The hot
nuclear system is modeled in a subtracted Thomas-Fermi framework. Compared
to semi-infinite nuclear matter, substantial changes in the 
surface symmetry energy of finite 
nuclei are indicated; surface thermodynamic
potential energies for the two systems are, however, not too different.
Analytic expressions to fit the temperature and asymmetry dependence
of the surface thermodynamic potential of semi-infinite nuclear
matter and the temperature dependence of the surface free energy 
of finite nuclei are given.
\end{abstract}



\maketitle

\section{Introduction}

The liquid-drop model provides a sound framework 
\cite{wei,mye1,mye2} for having a good estimate of the nuclear surface energy from experimental
binding energy systematics. This is the case for cold nuclei; this 
has helped, for instance in understanding barrier heights or saddle
point configurations in nuclear fission. This estimate 
has also its place in the
framing of effective nucleon-nucleon interactions \cite{mye1,ban1} by
providing an important empirical input. There is a strong motivation
too to examine the thermal properties of the nuclear surface. Hot
nuclei, produced in multifragmentation in nuclear collisions are
surrounded by nucleonic vapor, knowledge of the energy of the interface
between the nuclear liquid and vapor is a crucial determinant in
their mass distributions \cite{bon1,bon2} or in our understanding
of their thermodynamic limit of existence \cite{lev,ban2}. It has also 
important astrophysical applications. It puts significant constraints
in determining the equilibrium nuclear masses, electron capture
rates and level densities that play a seminal role in the dynamical
evolution of neutron stars and supernovae \cite{ste,jan}.   

Semi-infinite nuclear matter (SINM) offers a good starting ground for exploring
the nuclear surface properties. It has a simplicity coming from absence
of many undesirable complications arising from shell, Coulomb and  
finite-size effects. Considerable effort has been directed in the past
in understanding its surface properties at zero temperature, mostly
in the semi-classical Thomas-Fermi (TF) 
framework \cite{mye3,kol,cen}; studies
have also been done in the quantal Hartree-Fock approach \cite{pea,ton}. 
In a need for applications in astrophysical scenario, Ravenhall, Pethick
and Lattimer in their pioneering work \cite{rav} explored the thermodynamic
evolution of the surface properties of symmetric as well as asymmetric
nuclear matter in the semi-classical approach using a plausible Skyrme
interaction. With increasing temperature or asymmetry, the density of
vapor consisting of hot or drip nucleons surrounding the liquid phase
of the nuclear matter increases. The evolution of the interface energy
with this change was quantitatively evaluated by them; they showed 
how the surface thermodynamic potential energy or the surface entropy 
eventually dissolves at a critical temperature when the distinction 
between the nuclear liquid and vapor is lost. A temperature dependence 
of the surface thermodynamic potential 
energy of the form $[g(T,T_c(X))]^{\alpha_1}$ 
with $\alpha_1 =  $1.25 was suggested by them.
The function $g(T,T_c(X))$ has the form  
\begin{eqnarray}
g(T,T_c(X))= \left ((T_c^2(X)-T^2)/(T_c^2(X)+T^2) \right ), 
\end{eqnarray}
where $T_c(X)$ is the critical temperature for infinite nuclear matter
of isospin asymmetry $X$, defined as 
$X=(\rho_n -\rho_p)/(\rho_n+\rho_p)$ where $\rho_n$ and
$\rho_p$ are the neutron and proton densities
(for SINM, the definition of asymmetry is somewhat more subtle
and given later). Since
then, numerous calculations have been done to understand properties 
of hot nuclear matter \cite{bon1,bon2,de1} 
with this form of temperature-dependent
interface energy.

The scenario for finite nuclei is, however, different. 
The Coulomb interaction,
coupled with the microscopic nuclear size may influence the thermal
evolution of their surface differently from that of the semi-infinite
matter. It was already noticed, in the course of the evaluation of
the thermal dependence of volume and surface symmetry energy coefficients
\cite{de2} of nuclei with a Skyrme-type KDE0 interaction and the 
finite-range modified Seyler-Blanchard interaction, that the surface 
thermodynamic potential
of finite nuclear systems evolves in a slightly different way.
The form of the evolution function 
$h(T)=g(T,T_c(X=0))$ is the same, but the 
exponent $\alpha_1 $ is seen to have a 
different value, slightly different for 
the two interactions. 
The hot nuclei undergo
Coulomb instability at a limiting temperature \cite{lev}, which is 
much lower compared to the critical temperature $T_c$. Consequently,
the whole temperature range upto $T_c$ is not accessible for  
the finite nuclei. The limiting
temperature is generally a decreasing function 
of the atomic number \cite{ban1,ban2}.

On theoretical grounds, it is known  that for infinite systems the surface
thermodynamic energy behaves with temperature as $(T_c-T)^{\alpha_1} $,
with $\alpha_1 $=1.26 \cite{lan}. But that is near the critical point.
We note that for finite nuclei only a lower temperature range can 
be mapped. In this case, the value of the calculated exponent 
is found different \cite{de2}
from that found earlier in case of SINM.
We therefore intend to examine further in this 
article the evolution of the nuclear
interface energy with temperature for semi-infinite nuclear matter,
with a focus to the temperature range accessible to microscopic
nuclei. Calculations are done in the TF approximation. To assess 
any possible dependence of the  thermodynamic
surface energy on effective nucleon-nucleon interaction, three
Skyrme class interactions, namely, 
SkM$^*$ \cite{bar1}, SLy4 \cite{cha} and SK255 \cite{agr} are
employed in the calculations. These interactions 
obtained by accurately calibrating the bulk properties of finite nuclei
over the whole periodic table are quite successful.

Since the setting for  finite nuclear systems 
as stated earlier is  somewhat different,
calculations for their surface energies  are
also done with these interactions. Hot nuclei, because of
evaporation are inherently unstable. 
To give stability, the subtraction procedure was first suggested
by Bonche, Levit and Vautherin \cite{vau1,vau2} in the Hartree-Fock
framework. In the present work, we adopt its semi-classical variant, the
finite temperature Thomas-Fermi (FTTF) scheme \cite{sur}.
This ensures complete thermodynamic equilibrium between the high
density central liquid and the low density surrounding nucleon gas.
For a set of doubly closed shell and singly closed shell nuclei
covering almost the entire periodic table, the energies and free energies
are calculated in the FTTF scheme as a function of temperature. When
subjected to analysis in the framework of 
Bethe-Weizs\"acker liquid-drop mass formula, they
yield the temperature dependent surface thermodynamic entities.

The paper is organized as follows. Section II is devoted to 
the theoretical formulation of the study 
of the thermodynamic properties of the surface of asymmetric 
semi-infinite nuclear matter
and of finite nuclei. 
Results and discussions are presented in Sec. III. 
Conclusions are drawn in Sec. IV.

\section{The nuclear surface properties: the model}

Determination of  the equilibrium density distribution of the hot
nuclear systems is the starting point for calculations of the
thermodynamic properties of the nuclear surface. 
In order to describe a hot system as a stable one, it is assumed to be in
thermal equilibrium with a surrounding gas representing the evaporated
nucleons. Even a very asymmetric cold nuclear system may be stable 
beyond the nucleon drip point, the required stability being given
by the drip nucleons \cite{cen,de3}. The description of such a nuclear
liquid embedded in a gaseous environment can be given in FTTF framework.
Sec.~IIA describes the procedure for obtaining the equilibrium
density profiles for semi-infinite nuclear matter as well as for
finite nuclei. Sec.~IIB and IIC give a brief glimpse of how the different
surface properties are established from these density profiles.

\begin{table}[t]
\begin{center}
\caption{The values of the Skyrme parameters for SkM*, SLy4 and SK255
interactions.}

\begin{ruledtabular}
\begin{tabular}{lddd}
\multicolumn{1}{c}{Parameters}& 
\multicolumn{1}{c}{SkM*}&
\multicolumn{1}{c}{SLy4}&
\multicolumn{1}{c}{SK255}\\
\hline
$t_0$(MeV fm$^3$)&-2645.0&-2488.91&-1689.35 \\
$t_1$(MeV fm$^5$)&410.0&486.82&389.30  \\
$t_2$(MeV fm$^5$)&-135.0&-546.39&-126.07 \\
$t_3$(MeV fm$^{3(\gamma +1)}$)&15595.0&13777.0&10989.59 \\
$x_0$&0.09&0.834&-0.1461 \\
$x_1$&0.0&-0.344&0.116 \\
$x_2$&0.0&-1.0&0.0012 \\
$x_3$&0.0&1.354&-0.7449 \\
$\gamma $&0.1666&0.1666&0.3563 \\
\end{tabular}
\end{ruledtabular}
\label{data-gqr}
\end{center}
\end{table}
\subsection{Equilibrium density profiles}

The method  to obtain the equilibrium density profiles of 
semi-infinite matter and of finite systems is  based on 
the existence of two solutions to the TF
equations, one corresponding to the liquid phase with the surrounding
gas ($lg$) and the other corresponding to the gas ($g$) alone.
The two solutions are obtained from the variational equations
\begin{eqnarray}
\frac{\delta \Omega_{lg}}{\delta \rho_{lg}}=0,
\end{eqnarray}
and
\begin{eqnarray}
\frac{\delta \Omega_{g}}{\delta \rho_{g}}=0,
\end{eqnarray}
where $\Omega_{lg}$ and $\Omega_g$ are the thermodynamic potentials
of the said systems. These two systems have the same chemical potentials 
$\mu $ because of thermodynamic coexistence between the liquid plus gas
system and the embedding gas 
(i.e., $\mu_{lg}^q=
\mu_g^q=\mu_q, q$ refers to the isospin index for neutrons or protons). 
The base density profile of the
nuclear liquid $(l)$ of interest is obtained by subtracting the gas density
($g$) from that of the liquid plus gas system ($lg$), i.e., 
$\rho^q_l =\rho^q_{lg}-\rho^q_g $. 
 The thermodynamic potential is given by 
\begin{eqnarray}
\Omega =F-\sum_q\mu_qN_q,
\end{eqnarray}
where $F=E-TS$; $F, E$ and $S$ are the total free energy, energy and 
entropy, respectively, $T$ is the temperature and  $N_q$ the number
of neutrons or protons, $\mu_q$ being the corresponding chemical
potentials.

We have calculated the total energy with Skyrme interaction energy density
functionals. The energy density is 
\begin{eqnarray}
{\cal E }(r) = \frac{\hbar^2}{2m_n} \tau_n(r)+\frac{\hbar^2}{2m_p}
\tau_p(r) +{\cal E}_{sky}[\rho (r)] +{\cal E}_c (r),
\end{eqnarray}
where $\tau$'s are the kinetic energy density, ${\cal E}_{sky}$
is the interaction energy density and ${\cal E}_c$ is the Coulomb
energy density. The Skyrme interaction energy density is given by
\begin{eqnarray}
{\cal E }_{sky}[\rho (r)]&=
&\frac {1}{2}t_0[(1+\frac {1}{2}x_0)\rho^2-(x_0+\frac {1}{2})
(\rho_n^2+\rho_p^2)]  \nonumber \\
&&+\frac {1}{12}t_3\rho^\gamma [(1+\frac {x_3}{2})\rho^2
-(x_3+\frac {1}{2})(\rho_n^2+\rho_p^2)] \nonumber \\
&&+\frac{1}{4}[t_1(1+\frac{1}{2}x_1)+t_2(1+\frac{1}{2}x_2)]\tau \rho 
\nonumber \\
&&+\frac{1}{4}[t_2(x_2+\frac{1}{2})-t_1(x_1+\frac{1}{2})]
(\tau_n\rho_n+\tau_p\rho_p) \nonumber \\
&&+\frac {1}{16}[3t_1(1+\frac {1}{2}x_1)-t_2(1+\frac {1}{2}x_2)]
(\nabla \rho )^2 \nonumber \\
&&-\frac {1}{16}[3t_1(x_1+\frac {1}{2})+t_2(x_2+\frac {1}{2})]
[(\nabla \rho_n)^2+(\nabla \rho_p)^2], 
\end{eqnarray}
with $\tau=\tau_n+\tau_p$ and $\rho =\rho_n+\rho_p$. Here the $t_i$'s,
$x_i$'s and $\gamma$ are the Skyrme parameters listed in Table I, for
the three chosen interactions.
The Coulomb contribution is present in Eq.~(5) 
only for finite nuclei, they can not be treated
for both homogeneous nuclear matter and semi-infinite nuclear
matter. 

In Eq.~(6), we have not included the spin-gradient terms \cite{bra} as
they were ignored while fitting the parameters for the Skyrme forces
we have chosen.

At finite temperature, the effective kinetic energy density is \cite{bra}
\begin{eqnarray}
\tau_q^*=\frac {2m_{q}^*}{\hbar^2}A_{T,q}^*TJ_{3/2}(\eta_q),
\end{eqnarray}
with 
\begin{eqnarray}
A_{T,q}^*=\frac {1}{2\pi^2} \Bigl (\frac {2m_{q}^*T}{\hbar^2} \Bigr )^{3/2}.
\end{eqnarray}
In Eqs.~(7) and (8), $m_q^*$ is the nucleon effective mass,
\begin{eqnarray}
m_{q}^*&=&m\Bigl [1+\frac {m}{2\hbar^2} \Bigl  \{[t_1(1+
\frac {x_1}{2})+t_2(1+\frac {x_2}{2})] \rho \nonumber \\
&&+[t_2(x_2+\frac {1}{2})-t_1(x_1+\frac {1}{2})]\rho_q \Bigr \}\Bigr ]^{-1}, 
\end{eqnarray}
and $\eta_q$ the fugacity,
\begin{eqnarray}
\eta_q = (\mu_q -U_q)/T.
\end{eqnarray} 
In Eq.(10), $U_q$ is the nucleon single-particle potential,
\begin{eqnarray}
U_q = \frac{\delta ({\cal E}_{sky}[\rho (r)]+{\cal E}_c)}
{\delta \rho_q},
\end{eqnarray} 
the symbol $\delta $ referring to the functional derivative here.
For neutrons, ${\cal E}_c =0$.
The density $\rho_q$ is obtained as,
\begin{eqnarray}
\rho_q = A_{T,q}^* J_{1/2}(\eta_q).
\end{eqnarray} 
The $J_k$'s are the Fermi integrals. The $T$=0 case is a special case which
can be addressed easily from Eqs.(7)-(12).
The Coulomb energy density ${\cal E}_c(r)$ is the sum of the direct and
exchange contributions:
\begin{eqnarray}
{\cal E}_c(r)={\cal E}_c^D(r) + {\cal E}_c^{Ex}(r).
\end{eqnarray} 
The direct term is
\begin{eqnarray}
{\cal E}_c^D(r)=\pi e^2\frac{\rho_p(r)}{r}\int_0^\infty \rho_p(r^\prime )
[(r+r^\prime )-|r-r^\prime |]r^\prime dr^\prime ,
\end{eqnarray} 
and the exchange term is calculated from the Slater approximation as
\begin{eqnarray}
{\cal E}_c^{Ex}(r)=\frac{-3e^2}{4\pi }(3\pi^2)^{1/3}\rho_p^{4/3}(r).
\end{eqnarray} 
The entropy density ${\cal S}$ is the sum of contributions from neutrons
and protons:
\begin{eqnarray}
{\cal S}=\sum_q {\cal S}_q=\sum_q \Bigl (\frac{5}{3}\frac{\hbar^2}
{2m_q^*}\frac{\tau_q^*}{T}-\eta_q \rho_q \Bigr ).
\end{eqnarray} 

From Eqs.~(2) and (3), the coupled equations follow:
\begin{eqnarray}
T\eta^q_{lg}+U^q_{lg}=\mu_q,
\end{eqnarray}
\begin{eqnarray}
T\eta^q_{g}+U^q_{g}=\mu_q.
\end{eqnarray}
Solutions of the above two equations yield the required density profiles
$\rho^q_{lg}(r)$ and $\rho^q_{g}(r)$. The calculations proceed as follows:
from guess densities $\rho^q_{lg}(r)$ and $\rho^q_g(r)$, one calculates
$U_q$ and $\eta_q$ from Eqs.~(9)-(12), 
then obtains $\mu_q$ as
\begin{eqnarray}
\mu_q=\frac{1}{A_q}\Bigl \{\int [T\eta^q_{lg}+U^q_{lg}]\rho^q_{lg}
d^3r-\int [T\eta^q_g+U^q_g]\rho^q_g d^3r \Bigr \},
\end{eqnarray}
where $A_q$ is the neutron or proton number  of the subtracted liquid
part, 
\begin{eqnarray}
A_q=\int [\rho^q_{lg}(r)-\rho^q_g(r)]d^3r.
\end{eqnarray}
 With this $\mu_q$, one can calculate $\eta_q$ with the previous $U_q$ 
obtaining the next stage densities $\rho_q$ through Eq.~(12)
and proceed iteratively until convergence is achieved.

For a finite system, $A_q$ refers to the actual  number of 
neutrons $(N)$ or protons $(Z)$ in the
nucleus one is dealing with. For semi-infinite system, one chooses
for calculation a sufficiently large box size and suitable number of
nucleons so that both $\rho_g$ and $\rho_{lg}$ attain constant
values at large and short distances, respectively 
(the two extreme ends of the box as shown in  Fig.~1). 
The asymptotic constancy of $\rho_{lg}$ and $\rho_g$
at $z\rightarrow -\infty$ and $z\rightarrow +\infty$, respectively is
assured this way.

\begin{table}[t]
\begin{center}
\caption{The values of the parameters determining the surface
interface energy as a function of temperature and asymmetry for the
SkM$^*$, SLy4 and SK255 interactions.}
\begin{ruledtabular}
\begin{tabular}{lddd}
\multicolumn{1}{c}{Parameters}& 
\multicolumn{1}{c}{SkM*}&
\multicolumn{1}{c}{SLy4}&
\multicolumn{1}{c}{SK255}\\
\hline
$\sigma_{\mu}(0,0)$(MeV fm$^{-2}$)&1.055&1.135&1.060 \\
$C_0$&6.445&12.079&-4.558  \\
$\alpha_1$&0.916&0.898&0.914 \\
$\beta$&-0.184&-0.576&-0.344 \\
\end{tabular}
\end{ruledtabular}
\end{center}
\end{table}

\subsection {Calculation of surface energy coefficients: semi-infinite matter }

In the context of the subtraction method for isospin asymmetric
semi-infinite nuclear system, there could be two definitions \cite{mye3,cen}
of the nuclear interface energy. The definitions differ depending on 
whether one calculates the change in the total free energy or that in the
total thermodynamic potential of semi-infinite matter from the
corresponding quantities of the bulk matter. The subtraction of the constant
gas density from that of the liquid plus gas does not change the surface
profile of the semi-infinite matter (see Fig.~1). Delineating the 
liquid density $(l)$
one can define the surface free energy per unit area 
$\sigma_e$ \cite{mye3,cen} from 
\begin{eqnarray}
F_A=Af^l_B+\sigma_e {\cal A},
\end{eqnarray}
where $F_A$ is the total free energy of the semi-infinite liquid ($l$)
containing $A$ nucleons in a cylinder of area 
of cross-section ${\cal A}$ normal to
the liquid surface and $f^l_B$ is the free energy per particle of the
homogeneous bulk liquid . It may be pointed out that the actual 
calculations are performed over a finite range of $z$, 0$\le z \le
z_{max}$, (see Fig.~1) so that the total  number of nucleons $A$ in
the said cylinder in the liquid is finite.

From Eq.~(21), one gets for $\sigma_e$, 
\begin{eqnarray}
\sigma_e=\int_0^{z_{max}} dz \bigl [{\cal F}_{lg}(z)-{\cal F}_g(z)
\bigr ] -Af^l_B/{\cal A}.
\end{eqnarray}
Using the expression for $f^l_B$, one finds,
\begin{eqnarray}
\sigma_e&=&\int_0^{z_{max}} dz \Bigl \{\bigl ({\cal F}_{lg}(z)-{\cal F}_g(z)
\bigr ) -\bigl (\rho_{lg}(z)-\rho_g(z) \bigr )\nonumber\\
&&\frac{{\cal F}_{lg}^0-{\cal F}_g^0}{\rho_{lg}^0-\rho_g^0} \Bigr \},
\end{eqnarray}
where we have used Eq.~(20) for the evaluation of $A$.

In Eq.~(23), ${\cal F}_{lg}, {\cal F}_g $ etc. are the free energy
densities, the superscript $0$ referring to the corresponding bulk 
quantities. It may be noted that ${\cal F}_g(z)={\cal F}_g^0$
as the gas density is constant throughout. 

The surface thermodynamic potential per unit area $\sigma_{\mu}$
can likewise be defined from
\begin{eqnarray}
\Omega_A = \Omega^l_B + \sigma_{\mu}{\cal A},
\end{eqnarray}
whence 
\begin{eqnarray}
\sigma_{\mu}&=&\int_0^{z_{max}}dz\Bigl \{ {\cal F}_{lg}(z) -
{\cal F}_g^0 \nonumber \\
&&-\mu_n[\rho_{lg}^n(z)-\rho_g^n]-\mu_p[\rho_{lg}^p(z)-\rho_g^p]
\Bigr \}.
\end{eqnarray}
In Eq.~(24), $\Omega^l_B =-(P_{lg}^0-P_g^0){\cal A}z_{max}$ where
$P^0$ refers to the bulk pressure.
For thermodynamic equilibrium, the pressure $P_{lg}^0=P_g^0$,
hence the bulk thermodynamic potential of the liquid $(l)$ $\Omega_B^l$
=0. Further, $\mu_{lg}^n=\mu_g^n=\mu_n$ and
$\mu_{lg}^p=\mu_g^p=\mu_p$.

\begin{figure}
\includegraphics[width=0.9\columnwidth,angle=0,clip=true]{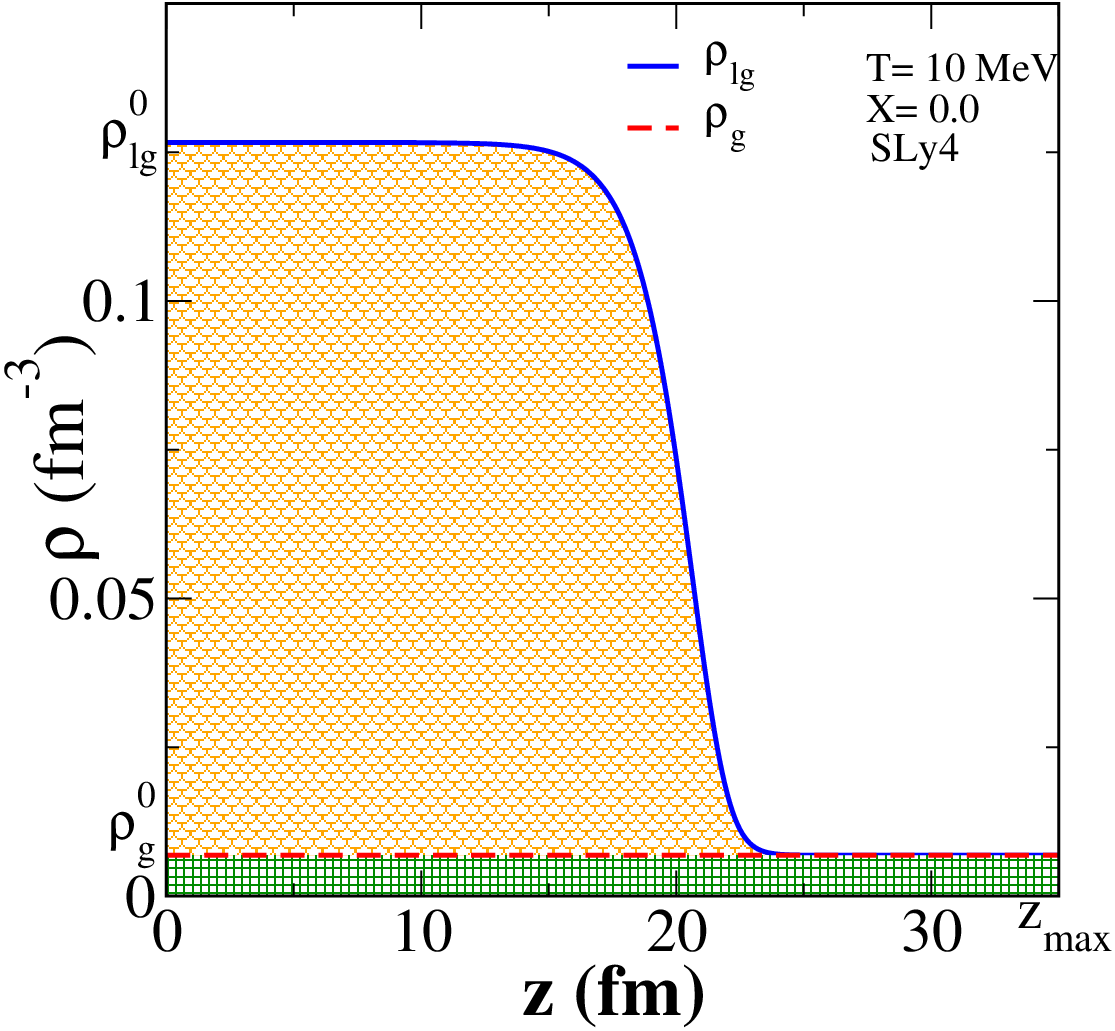}
\caption{(Color online) The density profiles for liquid plus gas 
($\rho_{lg}$) and gas ($\rho_g$) for symmetric semi-infinite
nuclear matter at $T$ =10 MeV with the SLy4 interaction. The green
shaded region and the purple shaded region represent the gas 
and the liquid density distributions, respectively. 
}
\end{figure}
\begin{figure}
\includegraphics[width=0.9\columnwidth,angle=0,clip=true]{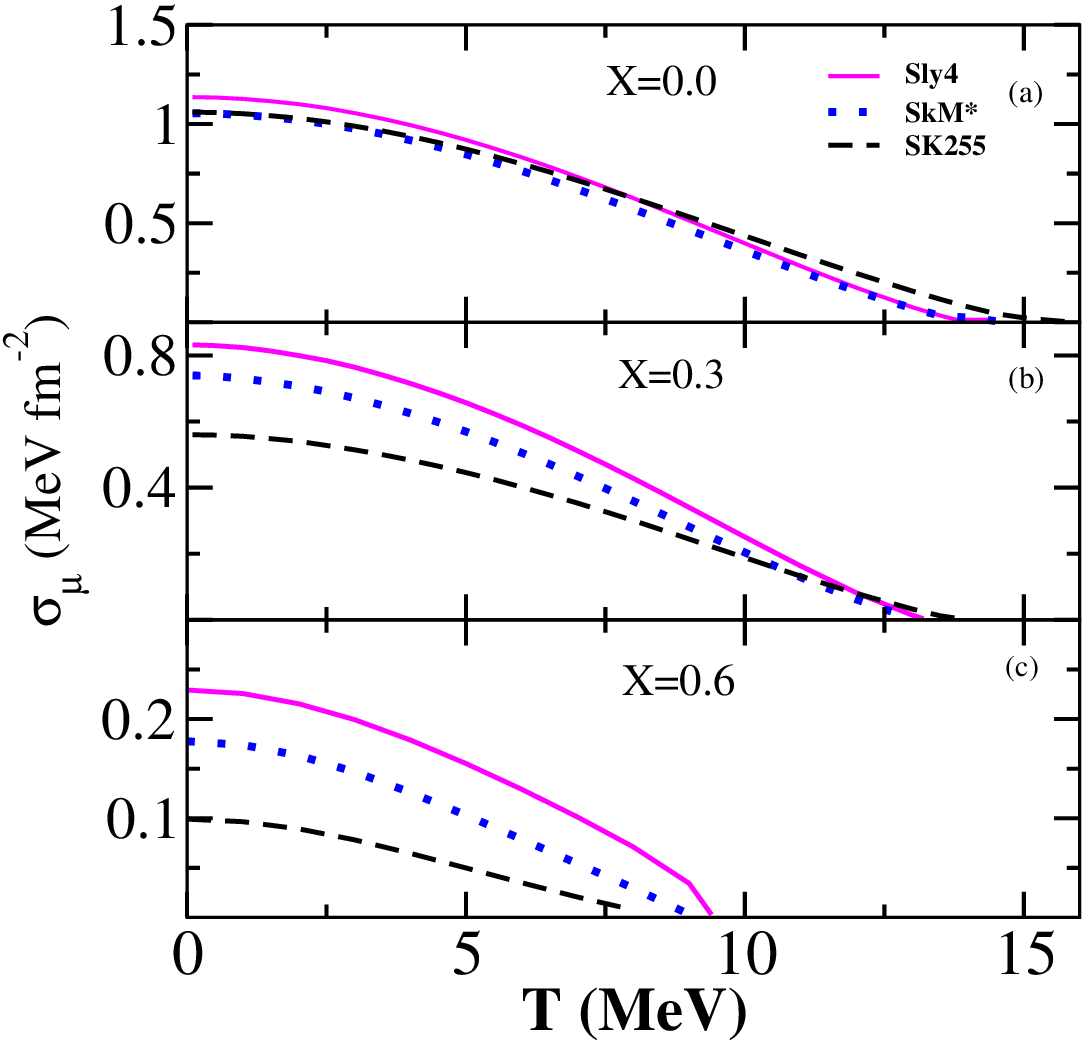}
\caption{(Color online) Thermal evolution of the surface thermodynamic
potential $\sigma_{\mu}$ of SINM with the SLy4, SkM$^*$ and SK255
interactions. Panels (a), (b) and (c) display results for $X=$ 0.0,
0.3 and 0.6, respectively.
}
\end{figure}
For isospin asymmetric systems, these two definitions given by 
Eqs.~(23) and (25) yield different surface interface energies.
The difference is given by \cite{cen}
\begin{eqnarray}
\sigma_e-\sigma_\mu =(\mu_n-\mu_p)(R_n-R_p)
\frac{\rho_{ln}^0\rho_{lp}^0}{\rho^0},
\end{eqnarray}
where $R_n$ and $R_p$ are the equivalent sharp surface locations
(in the spirit of the liquid-drop model) of the neutron and proton
fluid;  $\rho_{ln}^0,\rho_{lp}^0$ are the bulk  neutron
and proton densities in the liquid
and $\rho^0=\rho_{ln}^0+\rho_{lp}^0$. For 
symmetric matter, the two definitions yield identical results.

\begin{table}
\caption{Values of exponents $\alpha_1$ and $\alpha_2$ for symmetric SINM and
for finite nuclei in the temperature range T = 0 to 7.5 ($\sim T_{lim}$) MeV.}
\begin{tabular}{ccccccc}
\hline
\multicolumn{1}{c}{}&
\multicolumn{3}{c}{Symmetric SINM}&
\multicolumn{3}{c}{ Finite Nuclei}\\
\cline{2-7}
\multicolumn{1}{c}{Interaction}&
\multicolumn{1}{c}{$\alpha_1$}&
\multicolumn{1}{c}{$\alpha_2$}&
\multicolumn{1}{c}{$\alpha$}&
\multicolumn{1}{c}{$\alpha_1$}&
\multicolumn{1}{c}{$\alpha_2$}&
\multicolumn{1}{c}{$\alpha$}\\
\hline
SkM* & 0.966 & -0.076 & 0.814 & 1.486 & -0.222 & 1.042 \\
SLy4 & 0.919 & -0.069 & 0.781 & 1.404 & -0.226 & 0.952 \\
Sk255 &1.019  & -0.077 & 0.855 & 1.427 & -0.208 & 1.011 \\
\hline

\end{tabular}
\end{table}

\subsection {Calculation of surface energy coefficients: 
finite nuclei }

Unlike SINM, isospin asymmetry does not fully define 
the surface characteristics of atomic nuclei.
With the same isospin asymmetry, there may be nuclei 
with different neutron and proton numbers, 
the surface properties of nuclei may thus be
somewhat different. In that sense, one
can  talk meaningfully only about average surface properties of 
nuclei. To calculate the surface interface energy of hot finite
systems, we limit ourselves in the liquid-drop framework. In that
model, the total free energy of a nucleus is given by
\begin{eqnarray}
F(A,Z,T)&=&f_v(T)A+f_s(T)A^{2/3}+E_c(A,Z,T)   \nonumber \\
&&+(f_v^{sym}(T)-f_s^{sym}(T)A^{-1/3})AX^2 +....,
\end{eqnarray}
where $f_v$ and $f_s$ are the volume and the surface free energy 
coefficients for symmetric matter,
$E_c$ is the total Coulomb energy of the nucleus and $f_v^{sym}$
and $f_s^{sym}$ are the volume and surface free symmetry energy coefficients.
Here $X=(N-Z)/A$ is the isospin asymmetry of the nucleus. Referring to
our discussion in the previous subsection, one wonders 
whether ($f_s- f_s^{sym}X^2$) should be connected 
with $\sigma_e$ or $\sigma_{\mu}$.
It has been argued in Ref.~\cite{mye3} that in the context of the
liquid-drop model, it should be connected with $\sigma_{\mu}$,
the surface thermodynamic potential.
We follow the prescription. To calculate the thermal dependence of this
surface interface energy of finite nuclear systems, we evaluate,
in the subtraction scheme, the free energies of a host of spherical
or near-spherical nuclei, sixty nine in number (the list of nuclei
is taken from Ref. \cite{klu}), 
covering almost the 
entire periodic table ($34 \le A \le 218; 14 \le Z \le 92 $) at 
a finite temperature and make a least-squares fit of the calculated 
free energies with $f_v,~f_s$ etc. as free parameters. In actual
calculations, we fitted $F(A,Z,T)-E_c(A,Z,T)$, i.e., the nuclear
part of the free energy. The four parameters, namely, $f_v,~ f_s,~
f_v^{sym}$ and $f_s^{sym}$, so fitted, reflect  their desired 
temperature dependence.

\begin{figure}
\includegraphics[width=0.9\columnwidth,angle=0,clip=true]{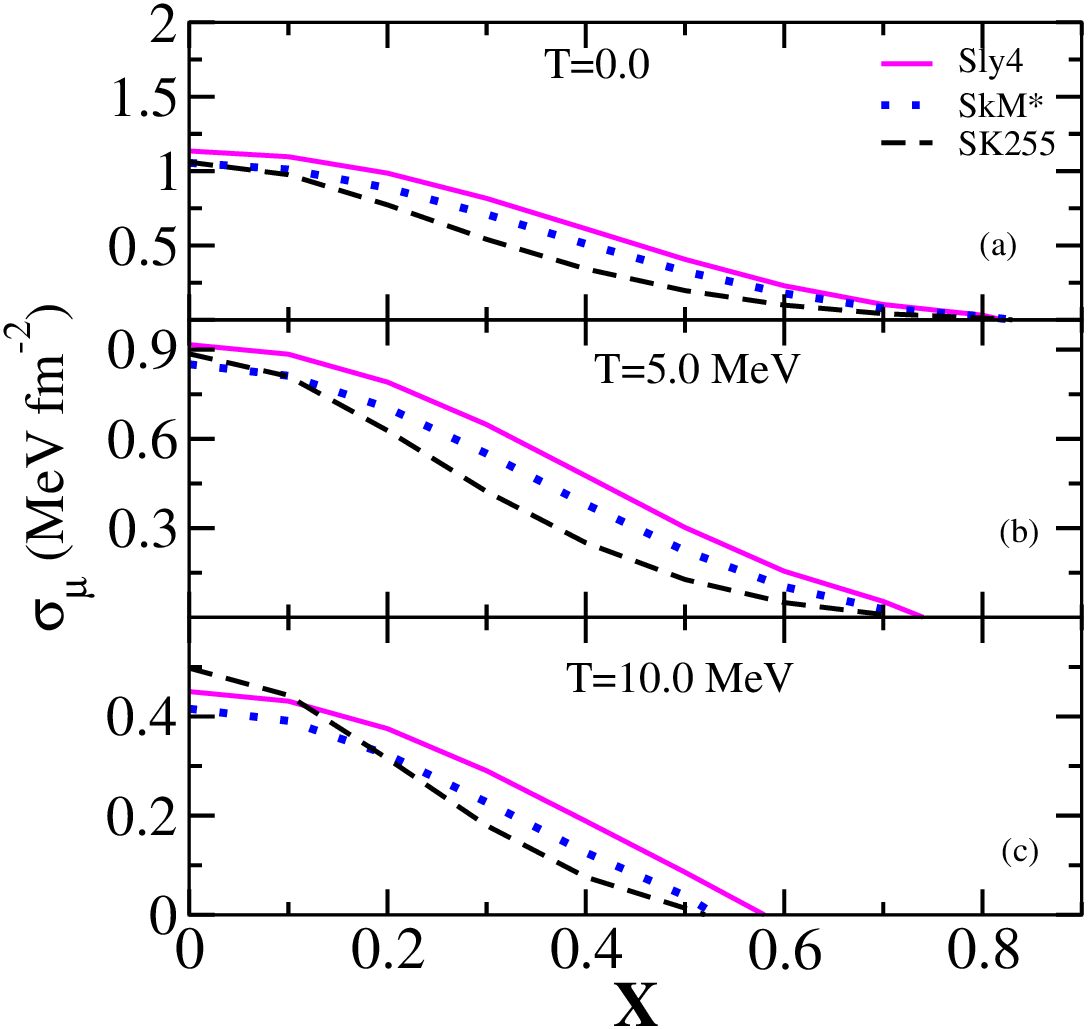}
\caption{(Color online) The surface thermodynamic potential $\sigma_{\mu}$
of SINM with the three interactions plotted as a function of 
isospin asymmetry. The three panels show results at temperatures
$T$ = 0.0, 5.0 and 10.0 MeV.
}
\end{figure}

\begin{figure}
\includegraphics[width=0.9\columnwidth,angle=0,clip=true]{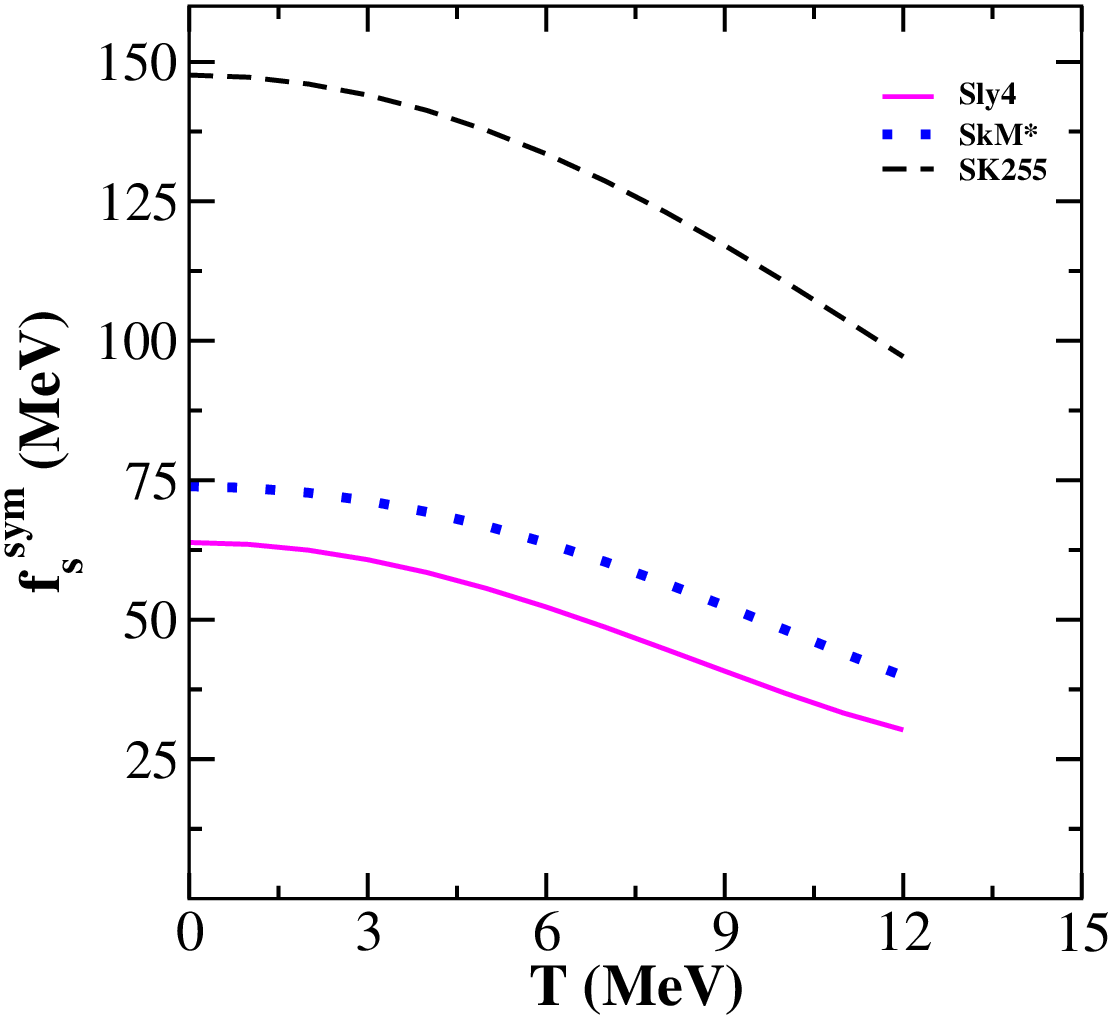}
\caption{(Color online) Thermal evolution of the surface symmetry free
energy coefficient $f_s^{sym}$ of SINM for the interactions SLy4,
SkM$^*$ and SK255.
}
\end{figure}

\section{Results and discussions}

The calculations for semi-infinite nuclear matter 
(SINM) and for finite nuclei
are done with three Skyrme interactions, SkM$^*$, SLy4 and SK255.
The actual calculations for SINM  are done in a reasonably
large box size. In Fig.~1, we display a typical density profile for
symmetric SINM at a temperature $T=$10 MeV with SLy4
interaction. As seen in the figure, with a box size of 35 fm, the gas
density on the right side of the box and the liquid plus gas density on
the left side attain constant asymptotic values. The liquid density
$\rho_l$ shown by the purple shaded region is obtained after subtracting
the constant gas density $\rho_g$ (shown by the green shaded region) from
the liquid plus gas density $\rho_{lg}$. The quantities $\rho_{lg}^0$
and $\rho_g^0$ represent the bulk values for the liquid plus gas and
gas densities, respectively. As seen in the figure, the choice of 35 fm 
for the maximum value of $z$ ($z_{max}$) suffices,  provided suitable
value for the number of nucleons per unit area is chosen (in the present
calculation, it is 2.5 per fm$^2$).
For asymmetric SINM, the neutron and proton fractions in the liquid
part and those in the gas part may be different. For the system in
phase equilibrium, definition of global asymmetry of the whole system
is then not practical, it depends on the size of the box. For numerical 
convenience, as in Refs.~\cite{rav,cen}, we therefore define asymmetry
as that of the denser side of SINM, i.e., 
$X=(\rho_{lg,n}^0-\rho_{lg,p}^0)/(\rho_{lg,n}^0+\rho_{lg,p}^0)$.

The thermal evolution of the surface thermodynamic potential per unit area 
$\sigma_{\mu}$ of semi-infinite nuclear matter is shown in Fig.~2 for
the three chosen interactions. The panels (a), (b) and (c) display calculated
results for different asymmetries. The general feature seen is that
$\sigma_{\mu}$ monotonically
decreases with temperature reaching zero at the critical temperature.
The value of the critical temperature depends on the choice of 
interactions and asymmetries.
 That $\sigma_{\mu}$
depends on the choice of interactions is evident. This dependence is
quite weak for symmetric matter,  becoming pronounced with
increasing asymmetry.

\begin{figure}
\includegraphics[width=0.9\columnwidth,angle=0,clip=true]{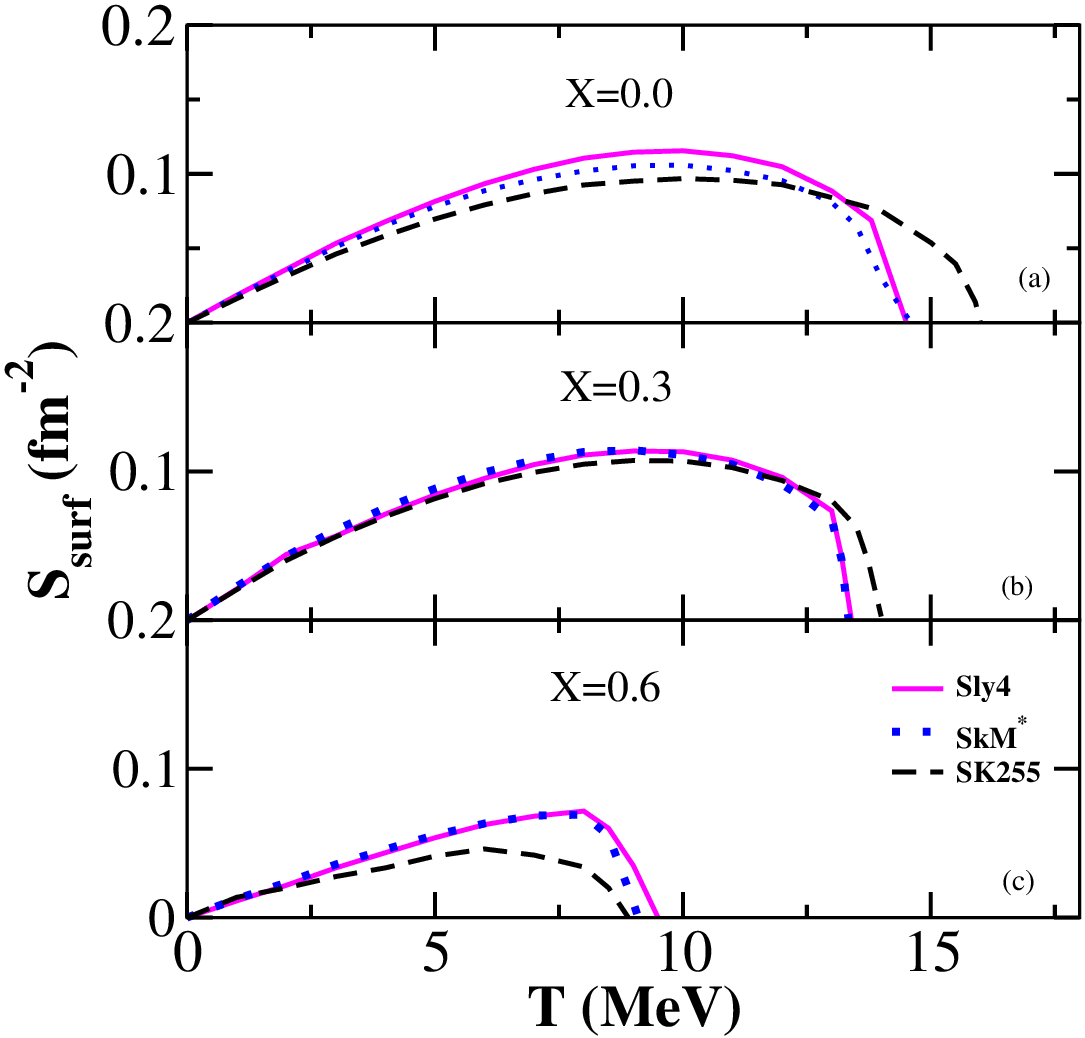}
\caption{(Color online) The surface entropy per unit area of SINM
shown as a function of temperature for the three interactions
at asymmetries $X$=0.0, 0.3 and 0.6.  
}
\end{figure}

\begin{figure}
\includegraphics[width=0.9\columnwidth,angle=0,clip=true]{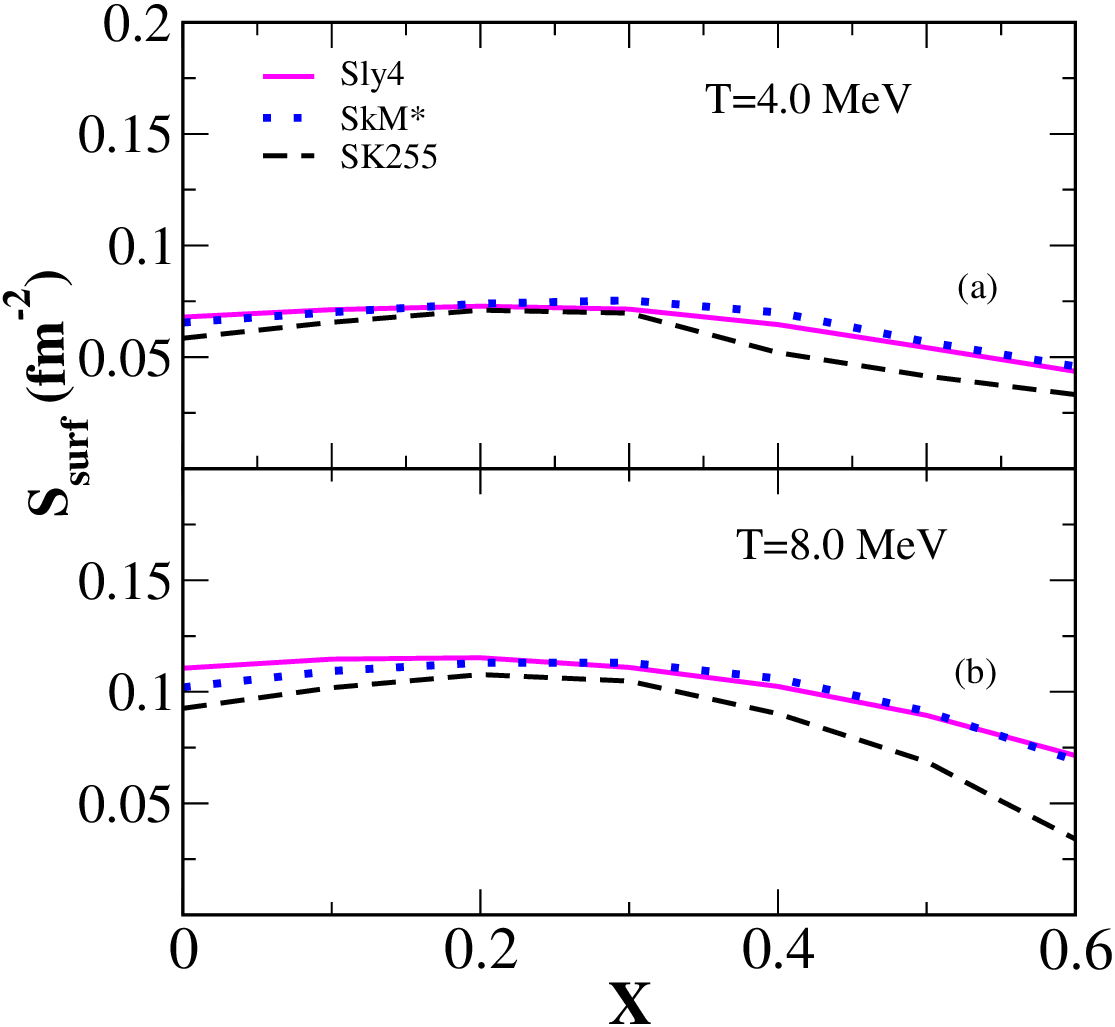}
\caption{(Color online) The surface entropy per unit area of SINM 
shown as a function of asymmetry for the three interactions at
$T$ =4.0 and 8.0 MeV.
}
\end{figure}
As shown in Ref.~\cite{lan}, near critical temperature, the surface
thermodynamic potential $\sigma_{\mu}(T,X)$ 
goes as $(T_c(X)-T)^{\alpha_1}$
with $\alpha_1 \simeq 1.26$. Keeping this in mind, the commonly used
temperature dependence of $\sigma_{\mu}(T,X)$ 
for SINM over the whole temperature 
range for all asymmetries has been taken as \cite{rav} 
\begin{eqnarray}
\sigma_{\mu}(T,X)=\sigma_{\mu}(0,X)[g(T,T_c(X))]^{\alpha_1},
\end{eqnarray}
where $g(T,T_c(X))$ is given by Eq.~(1) and $\alpha_1$=5/4. 
The asymmetry dependence of $\sigma_{\mu}(0,X)$ is taken as
\begin{eqnarray}
\sigma_{\mu}(0,X)=\sigma_{\mu}(0,0)(16+C_0)/[y^{-3}+C_0+(1-y)^{-3}],
\end{eqnarray}
where $y=(1-X)/2$ and $C_0$ a parameter.
The plausibility of the dependence of $\sigma_{\mu}(0,X)$
on $y^3$ has its origin on the phase equilibrium conditions \cite{rav}.
The critical temperature is asymmetry dependent.

From our calculation, we find that an algebraic expression of
$\sigma_{\mu}(T,X)$ of the form
\begin{eqnarray}
\sigma_{\mu}(T,X)&=&\sigma_{\mu}(0,0)[g(T,T_c(X))]^{\alpha_1}\nonumber\\
&& \times \frac {16+C_0[g(T,T_c(X))]^{\beta} } { y^{-3}
+C_0[g(T,T_c(X))]^{\beta}+(1-y)^{-3} },
\end{eqnarray}
gives an extremely good fit to the calculated values
for $X\leq 0.7$ and $T\leq T_c(X=0.7) \simeq 7.5$ MeV.
Upto a value of $X=0.7$, the asymmetry dependence of
$T_c$ can be well described by a polynomial of the form
$T_c(X)=T_c(X=0)[1+aX^2+bX^4]$. The values of $a$ are $-0.9238, -0.8126$
and $-1.2654$, the values of $b$ are -0.3529, -0.4193 and 0.1158,
for SkM$^*$, SLy4 and SK255 interactions, respectively. The values
of $\sigma_{\mu}(0,0), \alpha_1, C_0$ and $\beta $ are given in
Table~II for the three chosen interactions. The value of $\alpha_1$
is seen to be $\sim 0.9$ in this temperature range.
It has been checked that this value gradually rises
to the canonical value of $\simeq 1.26$ 
in a narrow temperature window near the critical temperature.
The critical temperatures $T_c(X=0)$ are
14.61, 14.53 and 15.98 MeV for SkM$^*$, SLy4 and SK255 interactions, 
respectively.  In Fig.~3, the asymmetry dependence of $\sigma_{\mu}$ is 
displayed for the same interactions at three temperatures, namely, at $T=$
0, 5 and 10 MeV in the the panels (a), (b) and (c), respectively. 
The surface potential 
$\sigma_{\mu}(T,X)$ decreases both with temperature and asymmetry
reaching zero at $T_c(X)$. One could see that
the temperature and asymmetry dependent $\sigma_{\mu}(T,X)$
has a non-negligible dependence on the
interactions one chooses to describe the semi-infinite
nuclear matter.

The external gas surrounding nuclear drops in clusterized nuclear
matter in astrophysical environment has its origin on both
temperature and asymmetry. Even at $T=0$, nuclei may exist 
embedded in a nucleon gas in astrophysical environment \cite{de3,papa}
leading to the modification of the nuclear properties. In Ref. \cite{papa},
Papakonstantinou {\it et al.} find an increase in surface energy
with increase in asymmetry for cold nuclei in contradiction to
what we find for SINM from Eq.~(30) with $T=0$. It may possibly be 
attributed to different definitions of the surface energy ($\sigma_e$
is known to increase initially with asymmetry \cite{cen}). Part of the 
reason may also lie in the definition of the asymmetry parameter;
$X$ in our case is the asymmetry of the denser part
of the liquid-gas system,
in \cite{papa}, the asymmetry defined is that of the subtracted
liquid part.

The temperature dependence of the surface symmetry energy coefficient
$f_s^{sym}$ for SINM is shown in Fig.~4.
In keeping with the convention used for finite systems (as is employed 
in Eq.~(27)), we  define $f_s^{sym}$ as
\begin{eqnarray} 
f_s^{sym}(T)=\Bigl [-4\pi r_0^2(T)\frac{1}{2}\frac {d^2\sigma_{\mu}(T,X)}
{dX^2} \Bigr ]_{X=0}
\end{eqnarray} 
where the radius parameter 
$r_0(T)=1/\bigl (\frac{4}{3}\pi \rho_l^0(T) \bigr )^{1/3}$,
$\rho_l^0 (T)$ being the bulk liquid density of symmetric matter
at the temperature concerned. From Eqs.~(30) and (31), one then gets
\begin{eqnarray} 
f_s^{sym}(T)&=&4\pi r_0^2(T)\sigma_{\mu}(0,0)g^
{\alpha -1}(T,T_c(X=0)) \nonumber \\
&&\times \Bigl [\frac{96g(T,T_c(X=0))}
{16+C_0g^{\beta}(T,T_c(X=0))}-\nonumber \\
&&\frac{4a\alpha T^2T_c^2(X=0)}{(T_c^2(X=0)+T^2)^2}\Bigr ].
\end{eqnarray} 
For all the interactions, the surface
symmetry energy decreases with temperature. 
It is evident from the figure that the calculated value of $f_s^{sym}(T=0)$
for SK255 interaction is around twice those obtained from  the other
two interactions. The comparatively faster fall of $\sigma_{\mu}$ with
asymmetry as seen in Fig.~2 for the SK255 interaction is a reflection of 
the larger value for the surface symmetry coefficient for this interaction.
The properties of symmetric and asymmetric nuclear matter at the
saturation density are quite different for the SK255 force compared
to the other two interactions used. For example, the nuclear 
incompressibility coefficient, the volume symmetry energy
coefficient and the density slope parameter $L$ 
($=3\rho \frac{\partial f_v^{sym}}{\partial \rho}$ at the saturation
density of symmetric nuclear matter at $T=0$) are larger
in comparison to those of the SkM$^*$ and SLy4 forces.
The large value of $f_s^{sym}(T=0)$ for SINM with SK255 
interaction can be qualitatively understood as being 
directly related to the corresponding large value of $L$
($\simeq $95 MeV for SK255 as compared to $\simeq $45 MeV
for SLy4 or SkM$^*$). For a heavy nucleus of mass number $A$,
$f_s^{sym}(T=0)$ is $\sim A^{1/3}[L\epsilon_A-\frac {K_{sym}}{2}
\epsilon_A^2]$ \cite{agr1}, where $K_{sym}$ is the symmetry
incompressibility and $\epsilon_A= (\rho_l^0-\rho_A)/(3\rho_l^0)$,
$\rho_A$ being the equivalent density for the nucleus. The quantity
$\rho_l^0-\rho_A$ can be parametrized as 
$\rho_l^0-\rho_A \simeq \rho_l^0/(1+cA^{1/3})$ \cite{cent} 
($c$ is around 0.28 for terrestrial  nuclei) so that for a 
hypothetical charge less large nuclear drop of mass $A$,
$\epsilon_A$ is $\sim A^{-1/3}/(3c) $. For SINM, $f_s^{sym} (T=0)$
is then $\sim L/(3c)$.

Entropy per unit area of the surface of semi-infinite
matter is discussed in association with the following two 
figures. The surface entropy per unit area is defined as \cite{rav}
\begin{eqnarray} 
s_{surf}&=&-\left. \frac{\partial \sigma_{\mu}(T,X)}{\partial T}\right|_{\mu_n}
\nonumber \\
&=&-\left.\frac{\partial \sigma_\mu (T,X)}{\partial T}\right|_X+
\left.\frac{\partial \sigma_\mu }{\partial X}\right|_T
\frac{\left.\frac{\partial \mu_n}{\partial T}\right|_X}{\left.\frac{\partial \mu_n}{\partial
X}\right|_T.}
\end{eqnarray} 
  In Fig.~5, the temperature dependence of $s_{surf}$ is shown for 
three asymmetries ($X=0.0, 0.3$ and $0.6$). The thermal evolution of
$s_{surf}$ for all the three interactions shows nearly the same
behavior for all asymmetries; temperature raises the
surface entropy as is expected, but after a maximum is reached, the
entropy falls sharply as the interface and the energy associated 
with it dissolve near the critical point. In Fig.~6, the asymmetry 
dependence of the surface entropy is displayed at two temperatures,
$T=$4.0 and 8.0 MeV. No marked sensitivity on either the interaction
or on $X$ except at large values of asymmetry is noticed.  

As discussed in Sec.~IIC, we have calculated the total free
energies of the nuclei as a function of temperature
in the subtracted FTTF procedure and then found the values
of the parameters $f_v(T), f_s(T), f_v^{sym}(T)$ and
$f_s^{sym}(T)$ from a least-squares fit.  
To be in concordance with the liquid-drop model of finite nuclei, a 
connection of $f_s(T)$ with $\sigma_{\mu} (T)$ ($\equiv \sigma_{\mu}(T,X=0)$) 
for symmetric semi-infinite 
nuclear matter is looked for, which can be established as 
\begin{eqnarray} 
f_s(T)=4\pi r_0^2(T)\sigma_{\mu }(T).
\end{eqnarray} 
For finite nuclei, $r_0(T)$ can be defined from $r_0(T)=R_0(T)/A^{1/3}$,
where $R_0(T)$ is its sharp surface radius. For semi-infinite
matter, $r_0(T)$  is already defined in connection with Eq.~(31). One can see
that $f_s(T=0) \sim $ 18 MeV, with $r_0(T=0) \sim $1.2 fm \cite{boh}.

The radius parameter $r_0(T)$ for both finite nuclei and symmetric
SINM is seen to be fitted extremely well in the temperature range 
$T=$0 to $T_{lim} (\sim 7.5)$ MeV as defined for finite nuclei
with the function $h(T)=g(T,T_c(X=0))$ as 
\begin{eqnarray} 
r_0(T)=r_0(0)[h(T)]^{\alpha_2},
\end{eqnarray} 
so that $f_s(T)$ can be defined as
\begin{eqnarray} 
f_s(T)=4\pi r_0^2(0)\sigma_\mu (0)[h(T)]^{\alpha},
\end{eqnarray} 
where $\alpha =\alpha_1 +2\alpha_2$. The values for $\alpha_1, \alpha_2 $
and $\alpha $ are listed in Table~III for symmetric SINM
as well as for finite nuclei for the three energy density functionals.
From the table, one sees that the exponent $\alpha_1$ governing the
temperature dependence of the interface energy per unit area
$\sigma_{\mu} (T)$ is significantly large for finite systems 
as compared to that of SINM. One further sees
that the radius parameter $r_0(T)$ increases faster for finite systems.
Overall, one finds that $f_s(T)$ for symmetric SINM and for finite
nuclei are not qualitatively very different though $f_s(T)$
seems to fall somewhat faster for finite nuclei.
In Fig.~7, this general thermal behavior of $f_s(T)$ is displayed.
The values of $\alpha_1, \alpha_2 $ and $\alpha $ are seen to be
nearly independent of energy density functionals. One may note
that there are some differences in the values of $\alpha_1$ for
SINM in Tables~II and III. This is so as $\alpha_1$ for SINM
in Table~III pertains to a small subset of data used for fitting
(symmetric SINM).
\begin{figure}
\includegraphics[width=0.9\columnwidth,angle=0,clip=true]{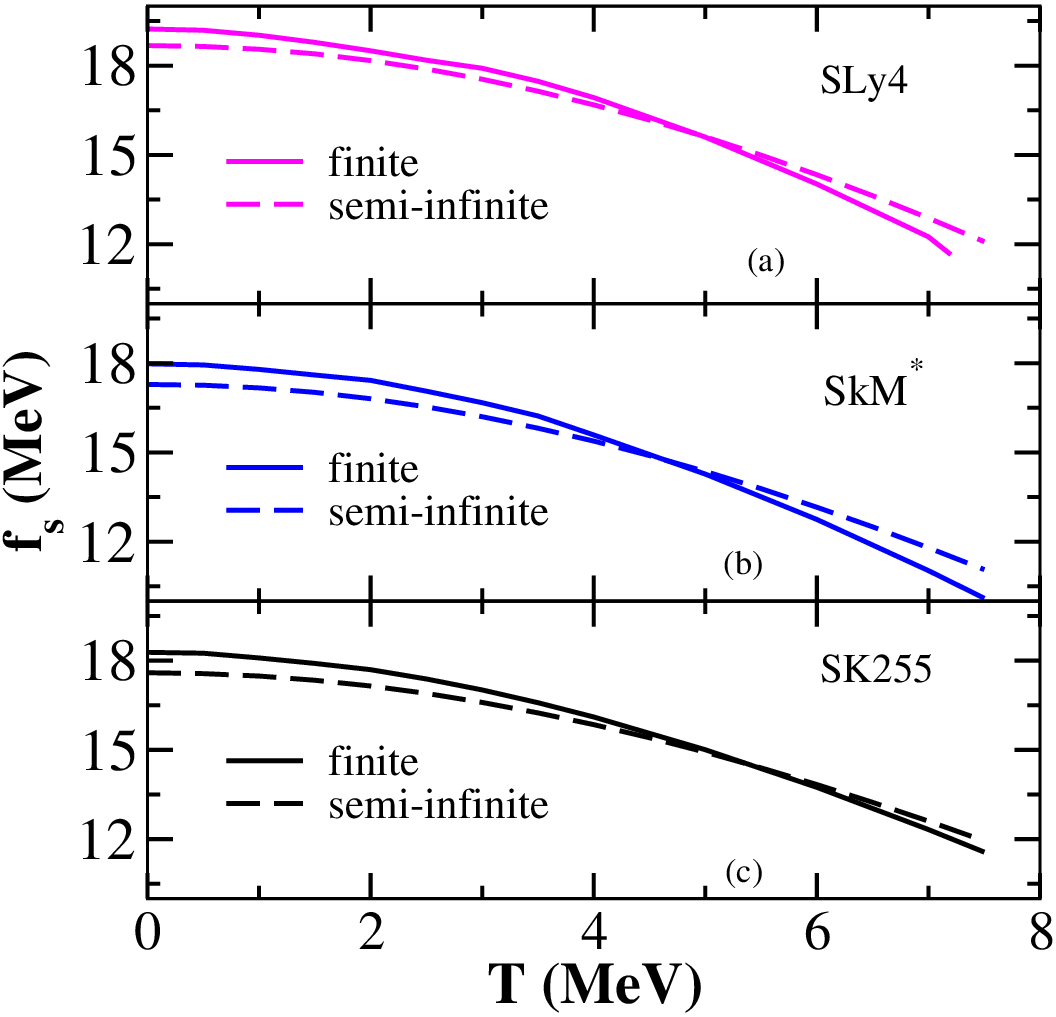}
\caption{(Color online) Comparison of the thermal evolution of the surface 
free energy $f_s$ for finite nuclei (full lines) and symmetric
SINM (dashed lines) with the interactions SLy4, SkM$^*$ and SK255.
}
\end{figure}
\begin{figure}
\includegraphics[width=0.9\columnwidth,angle=0,clip=true]{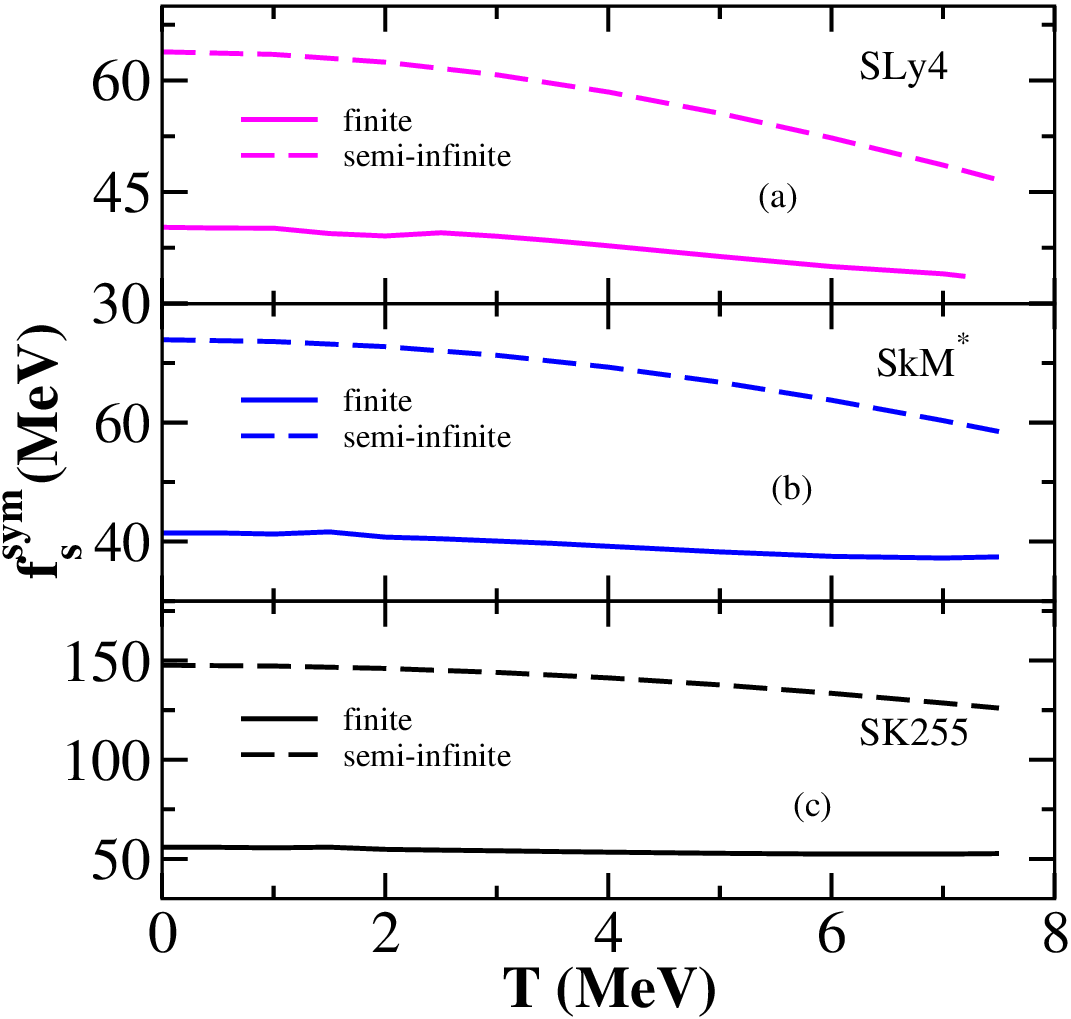}
\caption{(Color online) Comparison of the thermal evolution of the
surface symmetry free energy coefficients for finite nuclei (full
lines) and symmetric SINM (dashed lines) for the three interactions.
}
\end{figure}
The thermal evolution of the surface  symmetry free  energy coefficient 
$f_s^{sym}$ for finite nuclei is displayed for the three interactions 
in the panels (a), (b) and (c) in Fig.~8 and compared with that of
semi-infinite nuclear matter. Comparatively the sensitivity of 
$f_s^{sym}$ to temperature for finite nuclei is seen to be weaker.
Strikingly, $f_s^{sym}$ for semi-infinite matter is found to
be much larger for all the three interactions. This appears to
be a finite-size effect.
The Coulomb effect on $f_s^{sym}$ is found to be nominal.
Switching off the Coulomb interaction,
we have tested that as the nuclear size increases, $f_s^{sym}$
approaches  the asymptotic value for semi-infinite matter.
We take a set of nuclei at a fixed temperature (say $A \sim$500,
$T=0$) with different isospin asymmetries, calculate their free
energies with Coulomb switched off, and from a least-squares fit,
find the parameters $f_v, f_s, f_v^{sym}$ and $f_s^{sym}$
(cf. eq.~27) and then repeat the calculations at the same 
temperature for several other different sets of larger masses.
It is found that there is only a marginal change in
$f_v, f_s$ and $f_v^{sym}$, but $f_s^{sym}$ increases with the mass
of the nuclear set, tending asymptotically towards the SINM value.
The same conclusion emerges again even if $f_v, f_s$ and $f_v^{sym}$
are kept constant to the values specific for the interaction.
It may be noted that from double difference of {\it experimental }
symmetry energies of finite nuclei, the value of $f_s^{sym}(T=0)$
is empirically found to be 58.91$\pm $1.08 MeV \cite{jia}.
As is seen from Fig.~8, the value obtained from SK255 interaction 
is in consonance with this empirical value. Those from the other
two interactions are somewhat lower. The latter values, however, agree
closely with the value of $f_s^{sym}$ obtained from fitting of
nuclear masses \cite{sto}.

\section{Conclusions}

The thermal evolution of the surface properties of two-component
semi-infinite nuclear matter and of finite nuclei has been investigated
in the present article. Calculations are performed in the finite-temperature
Thomas-Fermi framework; stability to the seemingly unstable hot 
nuclear systems is achieved in the subtraction procedure. Three 
Skyrme-class interactions, namely SkM$^*$, SLy4 and SK255, designed
to reproduce the bulk properties of cold nuclei have been employed.
The dependence of the hot nuclear surface properties on the energy
density functionals are thereby explored.

 The combined effect of temperature and asymmetry on the nuclear
surface has important bearing  in astrophysics
and heavy ion collisions; in that context, for ready use, analytic
expressions that fit the calculated data for SINM well over a
wide range of temperatures and asymmetries are given.
For hot atomic nuclei, liquid-drop
model acted as a framework for obtaining the desired thermal evolution
of their surface. For applications in asymmetric systems, the need to
properly match the definition of the surface energy to the volume
energy has been stressed earlier \cite{mye3}. Due care has been taken 
in this work for its implementation in both finite nuclei and 
in semi-infinite nuclear matter; to be clear, propriety demands the 
evaluation of the surface thermodynamic potential which we have done.

The dependence of the surface thermodynamic potential on
temperature is seen to be of the form $[g(T,T_c(X))]^{\alpha_1}$;
for SINM, $\alpha_1$ has somewhat different values in different
temperature ranges, rising slowly from $\sim$ 1.0 at low temperature
to $\sim $ 1.26 near the 
critical temperature. For different interactions, $\alpha_1$
is seen to have nearly the same value. For finite nuclei, the functional
form of $g(T,T_c(X=0))$ remains the same, but $\alpha_1$ is much larger,
in the neighborhood of $\sim $1.45 for the three interactions that
we have chosen. These are finite size effects, they leave their
imprints on the surface symmetry free energy coefficient too; for
finite systems, the surface symmetry energy
 is comparatively much smaller, rising
slowly to the asymptotic value for semi-infinite nuclear matter
with increasing size. 

\begin{acknowledgments}
The authors gratefully acknowledge the assistance of Tanuja Agrawal
in the preparation of the manuscript.
J.N.D  acknowledges support from the 
Department of Science \&  Technology, Government of India.
\end{acknowledgments}

\end{document}